\documentclass[twocolumn,aps]{revtex4} %{revtex4-2}
\usepackage{mathrsfs}
\usepackage{amsmath}
\usepackage{graphicx}
\usepackage{dcolumn}
\usepackage{bm}
\usepackage{txfonts}
\usepackage{hyperref}
\usepackage{subfigure}
\usepackage{epstopdf}

\usepackage{xcolor,color}% coloured text
\usepackage{subfigure}
\usepackage{color}

\hypersetup{%
  breaklinks = {true},
  citecolor = {blue},
  colorlinks = {true},
  linkcolor = {red},
  pdfauthor = {\textcopyright\ Vitor M. Pereira},
  pdfcreator = {\LaTeX\ and \flqq hyperref\frqq},
  pdffitwindow = {true},
  pdfmenubar = {true},
  pdfpagelayout = {SinglePage},
  pdfstartview = {Fit},
  pdftoolbar = {true},
  plainpages = {false},
}

\begin{document}

\title{Van Hove singularity-induced negative magnetoresistance in Dirac semimetals}

\author{Kai-He Ding$^{1}$}
\email{dingkh@csust.edu.cn}

\author{Zhen-Gang Zhu$^{2,3,4}$}
\email{zgzhu@ucas.ac.cn}

\affiliation{$^{1}$ Department of Physics and Electronic Science, Changsha University of Science and Technology, Changsha 410076, China.\\
$^{2}$ School of Electronic, Electrical and Communication Engineering, University of Chinese Academy of Sciences, Beijing 100049, China. \\
$^{3}$ Theoretical Condensed Matter Physics and Computational Materials Physics Laboratory, College of Physical Sciences, University of Chinese Academy of Sciences, Beijing 100049, China.\\
$^{4}$ CAS Center for Excellence in Topological Quantum Computation, University of Chinese Academy of Sciences, Beijing 100190, China.}

\begin{abstract}
Negative magnetoresistance (NMR) is a marked feature of Dirac semimetals, and may be caused by multiple mechanisms, such as the chiral anomaly, the Zeeman energy, the quantum interference effect, and the orbital moment. Recently, an experiment on Dirac semimetal Cd$_3$As$_2$ thin films revealed a new NMR feature that depends strongly on the thickness of the sample [T. Schumann, \emph{et al}., Phys. Rev. B 95, 241113(R) (2017)]. Here, we introduce a new mechanism of inducing NMR via the presence of the van Hove singularity (VHS) in the density of states. Theoretical fitting of the experimental data on magnetoconductivity and magnetoresistance shows good agreement, indicating that the observed NMR in thin films of Cd$_3$As$_2$ can be attributed to the VHS. This work provides new insights into the underlying of Dirac semimetals.

\end{abstract}

%\pacs{68.65.-k, 72.15.Rn, 72.80.Ng}

%\pacs{ 73.43.-f, 72.80.Vp, 72.15.Rn, 78.67.Wj}

\maketitle

\textit{Introduction.---}
Negative magnetoresistance (NMR) is a phenomenon in which the electrical resistance of a material decreases when a magnetic field is applied. This effect has been observed in various materials \cite{pippard,beenakker,lin2011,mn2007,bloom2007}, including metals, semiconductors, and organic materials.
Recently, NMR has gained significant attention due to its observation in Dirac semimetals \cite{kimprl2013,tlnm2015,zhaosr2016,schumannprb2017,zklsci2014,xiongsci2015,linp2016,wanprb2011,xuprl2011,lvprx2015,aasnat2015,hxwprl2020}, a class of materials with a unique electronic band structure characterized by the presence of Dirac cones.
In Dirac semimetals, the NMR effect is particularly intriguing because it reveals fundamental properties of their electronic structure, such as the Berry phase \cite{xiaormp2000} and the chiral anomaly \cite{aazprb2012,dtsprl2012,dtsprb2013,ashby2014} by measurable physical quantities.
Consequently, a growing body of research has focused on understanding the NMR effect in Dirac semimetals theoretically and experimentally.
However, the origin of this behavior is not unveiled thoroughly to the bottom, and different mechanisms have been proposed,
% there is currently no consensus on the origin
%of this behavior. It is suggested that it may be caused by
including the chiral anomaly \cite{vaprb2012,dtsprl2012,ashby2014,dtsprb2013,aazprb2012,aabprl2014}, the Zeeman energy \cite{tlnm2015,xiongsci2015}, the quantum interference effect \cite{kimprl2013,xhprx2015}, the orbital moment \cite{daiprl2017}, or a highly non-uniform current distribution inside the sample \cite{reisnjp2016}.
Moreover, similar NMR behaviors have been experimentally reported in the Dirac system with non-zero mass \cite{wangnr2012,mutchsci2019}, which is believed to result from the quantum interference of Dirac fermions in the presence of strong coupling between the conduction and valence bands caused by scalar impurity scattering potentials \cite{fuprl2019}.
More recently, NMR has been observed in thin films of Cd$_{3}$As$_{2}$, which strongly depends on the thickness of the samples \cite{schumannprb2017}. As the thickness increases, the NMR changes into a positive V-shaped notch near zero magnetic field. This NMR feature cannot be well explained by previously proposed mechanisms, indicating buried physics of NMR.

In this letter, we propose a new mechanism for the observed NMR in Cd$_3$As$_2$ thin film which is named the van-Hove-singularity (VHS) induced NMR. %on the magnetoresistance of Dirac semimetals.
The presence of the VHS results in a higher density of states near the singular points, which in turn facilitates backscattering.
As a result, multi-channel competition caused by the VHS increases magnetoconductivity, leading to the NMR in Cd$_3$As$_2$ thin film. Our theoretical predictions align well with experimental data of magnetoconductivity and magnetoresistivity, indicating that the NMR observed in thin films of Cd$_3$As$_2$ can be attributed to the VHS.

%**********************************************************************************************************************************
\begin{figure}[tbp]
\centering
\includegraphics[width=0.42\textwidth]{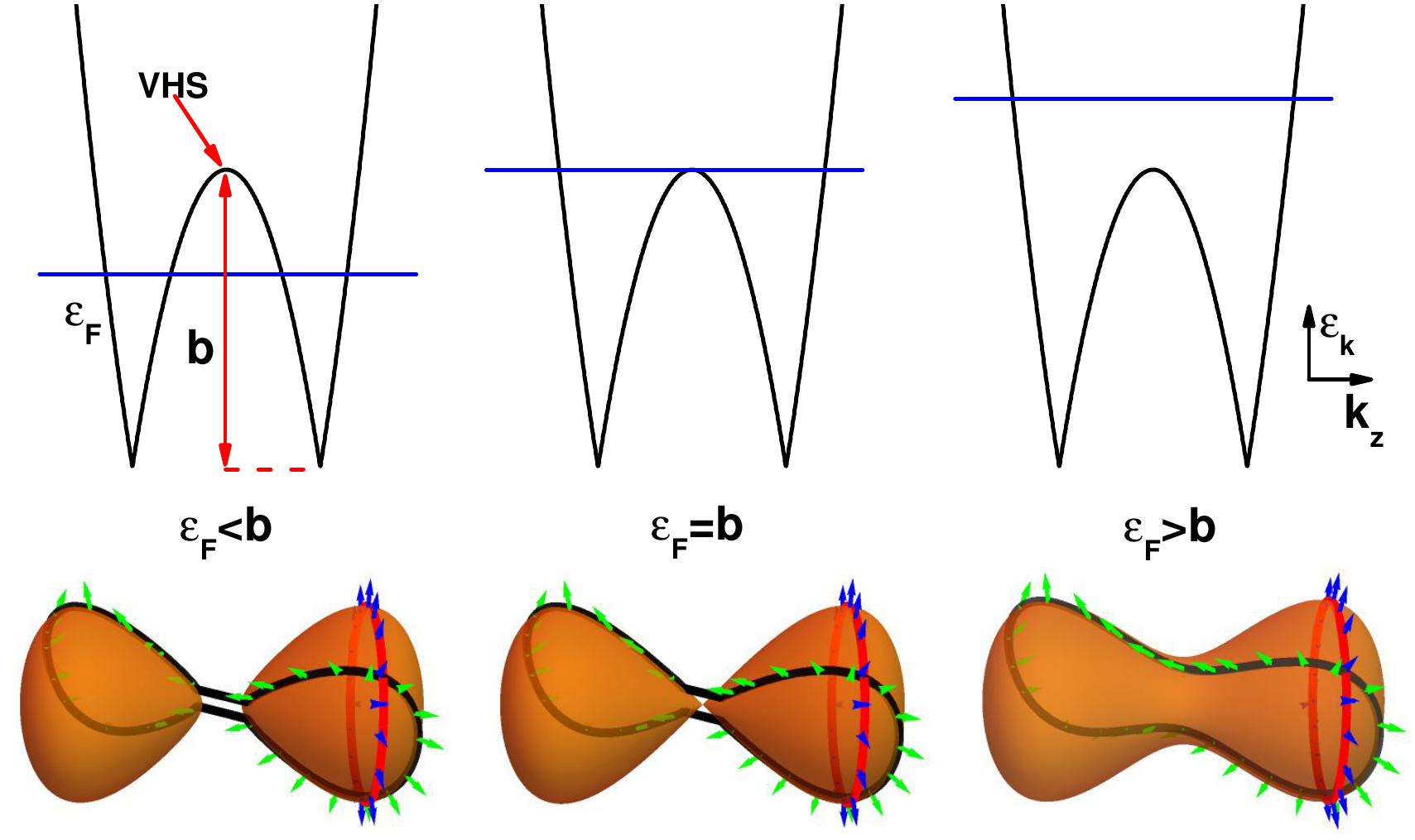}%0.38
\caption{Diagrams of the energy spectrum for the Dirac semimetal model of Eq. (\ref{rham1}), showing the crossing of the VHS by the Fermi level $\varepsilon_F$, i.e., from $\varepsilon_F<b$ to $\varepsilon_F>b$ via $\varepsilon_F=b$ at the fixed $b$ (Upper panels). The variation of the corresponding Fermi surface is depicted in the lower panels. The black and red curves on the Fermi surfaces, which enclose one and two Dirac cones, respectively, represent two distinct types of the scattering processes, where the green and blue arrows signify the spinor rotation occurring along the respective scattering paths. }\label{fig1}
\end{figure}
%************************************************************************************************************************************

\textit{Model.---}
Dirac semimetals are described by a low-energy effective Hamiltonian \cite{luprb2015,changprb2015,cortijoprb2016} as
\begin{equation}
H=\gamma( k_y\sigma_x- k_x\sigma_y)+\Delta\sigma_z,\label{rham1}
\end{equation}
where $k_i$ ($i=x,y,z$) are wave vectors, $\Delta=v_0k_z^2-b$, and $\sigma_i$ are the Pauli matrices. Solving the Schr\"{o}dinger's equation for Hamiltonian (\ref{rham1}), one can readily find the eigenvalue $\varepsilon_{\mathbf{k}\pm}=\pm\sqrt{\Delta^2+(\gamma k_\parallel)^2}$ and the corresponding eigenstates, $ |u^+\rangle=\left( \cos\frac{\theta_{\mathbf{k}}}{2}, -i\sin\frac{\theta_{\mathbf{k}}}{2}e^{i\phi_{\mathbf{k}}} \right)^T $, and $ |u^-\rangle=\left( \sin\frac{\theta_{\mathbf{k}}}{2}, i\cos\frac{\theta_{\mathbf{k}}}{2}e^{i\phi_{\mathbf{k}}}\right)^T $ with $k_\parallel=(k_x^2+k_y^2)^{1/2}$, $\tan\phi_{\mathbf{k}}=k_x/k_y$ and $\cos\theta_{\mathbf{k}}=\Delta/\varepsilon_{\mathbf{k}+}$. The band structure, schematically shown in Figure \ref{fig1}, consists of two Dirac cones separated by a saddle point at which $\nabla_{\mathbf{k}} \varepsilon_{\mathbf{k}}=0$, leading to the VHS.
Crossing the VHS, the constant energy contour undergoes a Lifshitz transition \cite{lifshitz1960,dingprb2021}, resulting in a topological change from two unconnected ellipsoid-like surfaces to a single surface (see Figure \ref{fig1}).
This VHS has been identified in Dirac semimetals using angle-resolved photoemission spectroscopy \cite{xuprb2015} and optical spectroscopy measurements \cite{chenprb2015}. The electron dynamics exhibits an unusual character near the VHS, which affects the properties of magnetotransport, which is the focus of this study.

Disorder is introduced by the Gaussian model \cite{Akkermansbook}. The average of the disorder potential is given by $\langle V(\mathbf{r})\rangle=0$, while the average of the two-point correlation function is $\langle V(\mathbf{r})V(\mathbf{r}')\rangle=n_iV_0^2 \delta(\mathbf{r}-\mathbf{r}')$, where $n_i$ and $V_0$ are the impurity density and strength, respectively. We assume that the Fermi level only intersects with the conduction band. Owing to the weak disorder potential, interband scattering is negligible \cite{Sinitsynprb2007}. %All calculations are limited to the conduction band.

%**********************************************************************************************************************************
\begin{figure}[tb]
\centering
\includegraphics[width=0.42\textwidth]{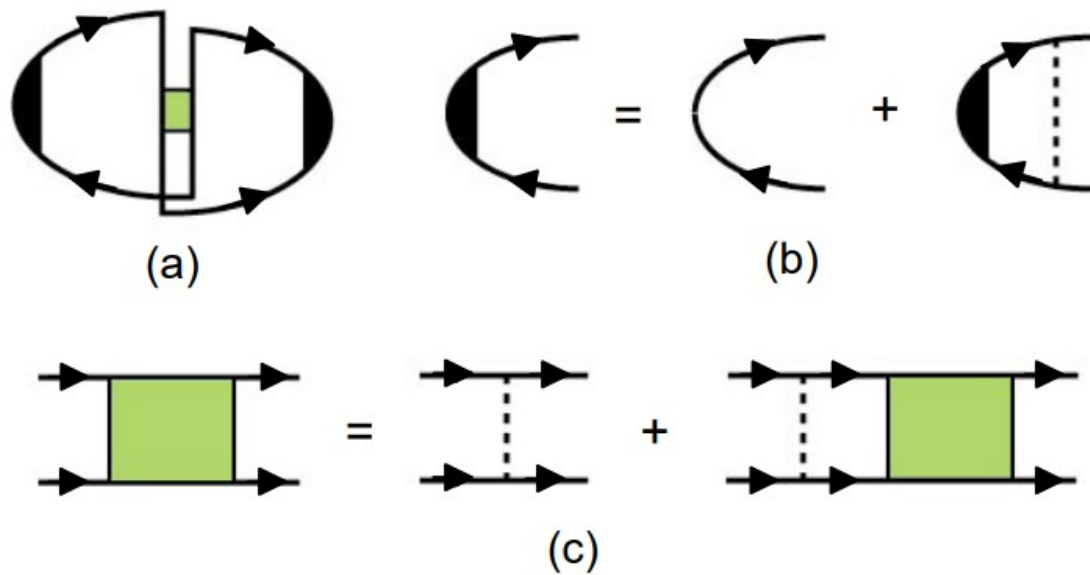}%0.38
\caption{(a) The Feynman Diagram for the conductivity correction. (b) The vertex correction to velocity. (c) The Bethe-Salpeter equation for Cooperon. The arrowed lines represent the impurity-averaged Green's functions, and the dash lines represent the impurity scattering. \label{hikami}}
\end{figure}
%************************************************************************************************************************************

\textit{Magnetoconductivity.---}
The conductivity correction can be computed using the Feynman diagrams called the Hikami boxes \cite{bruusbook,hikami1980, altshuler1980, mccann06}, shown in Figure \ref{hikami}(a), where the velocity's renormalization is obtained by the ladder diagrams in Figure \ref{hikami}(b).
Note that we ignore the dressed Hikami boxes because they make a small contribution. The calculation for the Hikami box is associated with the Cooperon, a two-particle correlation function, and obtained by solving the Bethe-Salpeter equation from the summation of the maximally crossed Feynman diagrams (see Figure \ref{hikami}(c)):
\begin{equation}
\Gamma_{\mathbf{k}\mathbf{k}'}(\mathbf{q})=\Gamma_{\mathbf{k}\mathbf{k}'}^0 +\frac{1}{V}\sum\limits_{\mathbf{k}_1}\Gamma^0_{\mathbf{k}\mathbf{k}_1} G_{\mathbf{k}_1}^R(\varepsilon) G_{\mathbf{q}-\mathbf{k}_1}^A(\varepsilon)\Gamma_{\mathbf{k}_1\mathbf{k}'}(\mathbf{q}), \label{bs0}
\end{equation}
where $\mathbf{q}=\mathbf{k}+\mathbf{k}'$, $V$ is the volume, and the retarded (advanced) Green's function is expressed as $G_{\mathbf{k}}^{R(A)}(\varepsilon)=1/(\varepsilon-\varepsilon_{\mathbf{k}+}\pm i/2\tau)$. In the Born's approximation, the impurity scattering time is given by $1/\tau=\frac{1}{\tau_0}\left(1+\frac{\Delta\langle \Delta\rangle}{\varepsilon_F^2}\right)$, where $\tau_0=\pi n_iV_0^2\rho(\varepsilon_F)$, and $\rho(\varepsilon_F)$ is the density of state at the Fermi level. $\langle \Delta\rangle$ is the average over the Fermi surface $\langle \Delta\rangle=\frac{1}{\rho(\varepsilon_F)}\sum\limits_{\mathbf{k}}\Delta\delta(\varepsilon_F-\varepsilon_{\mathbf{k}+})$. Further calculation shows $\langle\Delta\rangle=\frac{1}{3}[(b^2-\varepsilon_F^2)^{1/2}-b]$ at $\varepsilon_F<b$ and $\langle\Delta\rangle=\frac{1}{3}(\varepsilon_F-2b)$ at $\varepsilon_F>|b|$. Evidently, $\langle\Delta\rangle$, depending on the Fermi energy, decreases at $\varepsilon_F<b$, while it rises at $\varepsilon_F>b$. The transport-related quantities will be expressed in terms of $\langle\Delta\rangle$, giving rise to a different effect from the case of topological insulators, where $\Delta$ is a constant \cite{luprl2011}.

 The Cooperon, denoted by $\Gamma_{\mathbf{k}\mathbf{k}'}(\mathbf{q})$, encapsulates the process of multiple scattering along time-reversed paths, with its significance pronounced in contributions from small $\mathbf{q}$ values near zero.
In contrast to the treatment method in the spin space\cite{tkachovprb2011}, within the eigenstate framework, the bare vertex $\Gamma_{\mathbf{k}\mathbf{k}'}^0$ depends on the angles of the momenta, designated as $\phi_{\mathbf{k}}$ and $\phi_{\mathbf{k}'}$, persisting even after undergoing an impurity averaging procedure (refer to Eq. (13) in the supplemental material\cite{supp}). Through iterative analysis, we establish that the Cooperon takes the form $\sum_{nm\in 0,1,2}A_{nm}e^{i(n\phi_{\mathbf{k}'}-m\phi_{\mathbf{k}})}$, where $A_{nm}$ denotes expansion coefficients independent of the angles $\phi_{\mathbf{k}'}$ and $\phi_{\mathbf{k}}$ (see Equation (14) in the supplementary material\cite{supp}). Substituting this expression into the Bethe Salpeter equation and expanding the advanced Green's function $G_{\mathbf{q}-\mathbf{k}_1}^A(\varepsilon)$ up to the second order in $q^2$, we derive, while retaining the most divergent terms in the limit of $\mathbf{q}\rightarrow0$,
\begin{equation}
\begin{array}{cll}
\Gamma_{\mathbf{k}\mathbf{k}'}(\mathbf{q})&=&
\frac{\lambda_1\cos^2\frac{\theta_{\mathbf{k}}}{2}
\cos^2\frac{\theta_{\mathbf{k}'}}{2}}{X_1+q_{\parallel}^2+D_{1}q_z^2}
+\frac{\lambda_2\sin\theta_{\mathbf{k}}\sin\theta_{\mathbf{k}'}}{X_2+q_{\parallel}^2+D_{2}q_z^2}e^{i(\phi_{\mathbf{k}'}-\phi_{\mathbf{k}})}
\\
&&+\frac{\lambda_3\sin^2\frac{\theta_{\mathbf{k}}}{2}
\sin^2\frac{\theta_{\mathbf{k}'}}{2}}{X_3+q_{\parallel}^2+D_{3}q_z^2}
e^{i2(\phi_{\mathbf{k}'}-\phi_{\mathbf{k}})},
\end{array}\label{ckkco1}
\end{equation}
where $q_{\parallel}^2=q_x^2+q_y^2$.
From Eq.\eqref{ckkco1}, one notices that the Cooperon displays specific characteristics that depend on the individual momenta $\mathbf{k}$ and $\mathbf{k}'$. Furthermore, Eq.(\ref{ckkco1}) includes three channels.
These characteristic features emanate from the eigenstate representation, a formalism that sharply diverges from the outcomes established in the spin space and is similarly manifested in two-dimensional Dirac systems (see, e.g., \cite{luprl2011, suzuura02, haizhou2013}).
 Additionally, each channel in Eq. (\ref{ckkco1}) is characterized by the parameter $X_i$. The parameter $X_i$ acts as a gap in each channel, hindering the diffusion of electrons by modifying the mean free path and phase coherence length \cite{daiprb2016, chenprl2019}. Notably, when the energy of the VHS is assumed to be infinite, channels 1 and 3 disappear, and Eq. (\ref{ckkco1}) simplifies to the results of the single-node model with only one channel \cite{suzuura02,daiprb2016}. In this case, the presence of multiple channels necessitates opening a gap between the conduction and valence bands through magnetic doping or other methods \cite{fuprl2019, luprl2011}.

Solving the Feynman diagram in Figure \ref{hikami}(a), we obtain the conductivity correction that involves an integration over $\mathbf{q}$. To perform the integration, we introduce a cutoff in the form of two circles with radii $1/l_e$ and $1/l_\phi$, i.e., $1/l_e < q < 1/l_\phi$, where $l_e$ and $l_\phi$ represent the mean free path and the phase coherence length, respectively. The Cooperon plays a pivotal role in shaping the expression for the conductivity correction. It incorporates the phase factors $e^{i(\phi_{\mathbf{k}'}-\phi_{\mathbf{k}})}$ and $e^{i2(\phi_{\mathbf{k}'}-\phi_{\mathbf{k}})}$. As $\mathbf{q}$ approaches zero, $e^{i(\phi_{\mathbf{k}'}-\phi_{\mathbf{k}})}$ equals $e^{i\pi}=-1$ and $e^{i(\phi_{\mathbf{k}'}-\phi_{\mathbf{k}})}$ equals $e^{i2\pi}=1$.  By substituting this Cooperon expression into the conductivity correction, we then derive the conductivity correction at zero magnetic field as
\begin{equation}
\sigma_0
=\frac{e^2\lambda^2}{\pi \hbar}\sum\limits_{i=1}^3(-1)^i
w_{i}\mathcal{T}_{i}.
\label{sigeven}
\end{equation}
with the detailed derivation presented in Ref. \cite{supp}.
It is emphasized that within the spin-dependent framework, the computation of quantum corrections involves the summation of inverses of Cooperon eigenvalues acquired in the Hilbert space of spin momentum. Notably, these inverses of Cooperon eigenvalues encompass contributions from interband scattering\cite{fuprl2019}. In this context, Eq. \eqref{sigeven} is formulated within the eigenstate basis, as elucidated in Eq. (24) of the supplementary material\cite{supp}. Importantly, we assume the presence of a weak disorder potential, enabling the exclusion of interband scattering and facilitating Cooperon calculations exclusively within the conduction band. As a result, our findings distinguish themselves from the conventional formula for quantum corrections, typically established in the Hilbert space of spin momentum.
In Eq.\eqref{sigeven}, where the factor $(-1)^i$ arises from the combination of the phase factors $e^{i(\phi_{\mathbf{k}'}-\phi_{\mathbf{k}})}$ and $e^{i2(\phi_{\mathbf{k}'}-\phi_{\mathbf{k}})}$ within the Cooperon, along with the negation of the conductivity expressions.
When an external magnetic field is applied, we replace $\mathbf{q}$ in the Cooperon with the Landau level to account for the effect of the magnetic field. As a result, the $\mathbf{q}$ integral in the conductivity expression becomes a sum over the Landau levels, which yields the conductivity correction for the system
\begin{equation}
\begin{array}{cll}
\sigma_B
&=&\frac{e^2\lambda^2}{2\pi^2 h}
\displaystyle{\sum\limits_i} (-1)^i w_{i} \left\{-\displaystyle{\int_{\frac{1}{l_{\phi}}}^{\frac{1}{l_{e}}}dq_z}\psi\left(D_{i}l_B^2q_z^2+l_B^2X_i+\frac{1}{2}\right) \right. \\
&+&\displaystyle{\int_{0}^{\frac{1}{l_{e}}}dq_z} \psi\left((D_{i}-1)l_B^2q_z^2+l_B^2X_i+\frac{l_B^2}{l_e^2}+\frac{1}{2}\right) \\
&-& \left.\displaystyle{\int_{0}^{\frac{1}{l_{\phi}}}dq_z}\psi\left((D_{i}-1)l_B^2q_z^2+l_B^2X_i+\frac{l_B^2}{l_\phi^2}+\frac{1}{2}\right) \right\},\\
%&&\right\},
 \end{array}\label{msigeven1}
\end{equation}
where $\psi(x)$ is the digamma function, and $l_B=(\hbar/4eB)^{1/2}$ is the magnetic length. The magnetoconductivity is defined by $\delta\sigma(B)=\sigma_B-\sigma_0$, and further expressed as (see Ref. \cite{supp})
\begin{equation}
\delta\sigma(B)=
\sum\limits_iw_{i}\delta T_i,
\label{magetoc}
\end{equation}
where $\delta T_i$ stands for the contribution from the $i$-th Cooperon channel. In contrast to previous results \cite{suzuura02,daiprb2016}, where each channel consists of only one term, each $\delta T_i$ in this work contains three terms that arise from the anisotropy of the band structure (see Eq. (\ref{msigeven1})). This yields a competition within each channel, resulting in a more complex behavior of the magnetoconductivity.

%**********************************************************************************************************************************
\begin{figure}[tbp]
\centering
\includegraphics[width=0.45\textwidth]{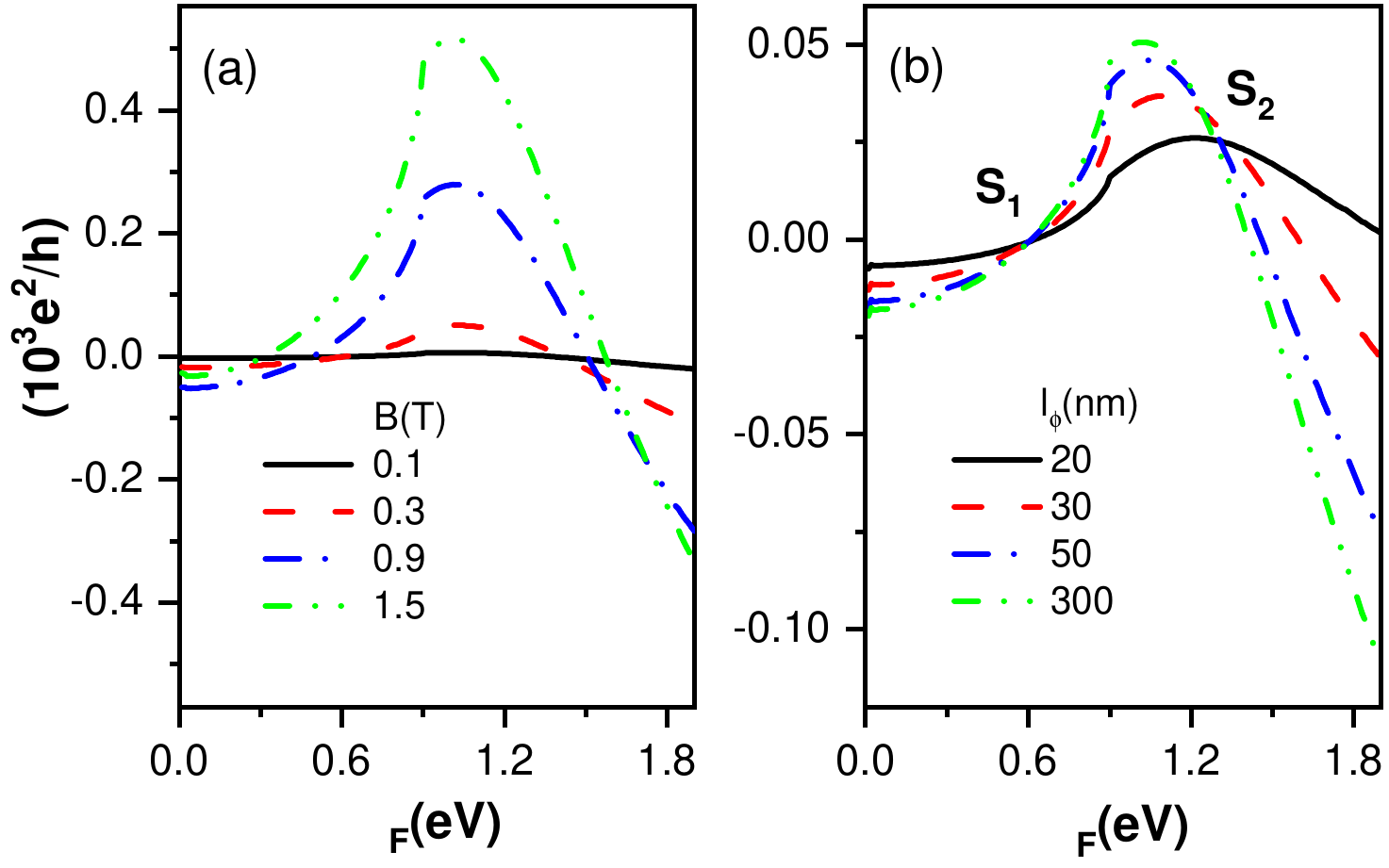}%0.38
\caption{Magnetoconductivity as a function of the Fermi energy $\varepsilon_F$ for the different magnetic field $B$ at $l_\phi=300$nm (a), and for the different phase coherence length $l_{\phi}$ at $B=0.3$T(b). %The inset shows the quantum correction to the conductivity versus the Fermi energy at the zero field and $l_\phi=30nm$.
%The dimensionless weight factors $w_i$ and the contribution of the Cooperon channel to the magnetoconductivity(in units of $10^3e^2/h$)  are shown in (c) and (d), %respectively. % Magnetoconductivity versus the magnetic field for the different Fermi energy $\varepsilon_F$ at $l_\phi=100nm$ (e).
The other parameters are taken as $\gamma=1.0$ eV$\cdot$nm, $v_0=0.6$ eV$\cdot$nm$^2$, $b=0.9$eV, and $l_e=10$nm.\label{fig3}}
\end{figure}
%************************************************************************************************************************************
Figure \ref{fig3}(a) illustrates the magnetoconductivity as a function of the Fermi energy for various magnetic fields. A prominent peak consistently emerges in the magnetoconductivity at the VHS. Although its height decreases as the magnetic field strength decreases, the amplitude still surpasses $\delta\sigma(B)=0$ even at very low magnetic fields. Importantly, this persistent $\delta\sigma(B)>0$ feature withstands modifications to other intrinsic parameters.We examine the influence of phase coherence lengths on the magnetoconductivity at weak magnetic fields in Figure \ref{fig3}(b). We observe that the magnetoconductivity curves intersect near two points, namely $S_1$ and $S_2$. $S_1$ corresponds to $\delta\sigma(B)=0$, while $S_2$ represents a value above $\delta\sigma(B)=0$. As the phase coherence length increases, the positions of $S_1$ and $S_2$ remain nearly unchanged, but the magnetoconductivity curve becomes steeper in the vicinity of these points. Consequently, the region between $S_1$ and $S_2$, where $\delta\sigma(B)>0$, is consistently maintained.

We can understand this characteristic behavior from the perspective of the Berry phase. Under the influence of the impurity potential, two types of scattering paths exist (see Figure \ref{fig1}).
When an electron follows the path surrounding two Dirac cones (path 1, indicated by the black curve in lower panel of Figure \ref{fig1}), its spin initially rotates counterclockwise, then switches to clockwise, and eventually returns to counterclockwise rotation. The net spin rotation throughout this process is zero, resulting in a zero Berry phase. Consequently, interference in the scattering process along path 1 is constructive, supporting positive magnetoconductivity. Conversely, when the electron follows the path surrounding one Dirac cone (path 2, denoted by the red curve in lower panel of Figure \ref{fig1}), the net spin rotation is $2\pi$, yielding a Berry phase of $\pi$ for the electrons. This leads to destructive interference in the backscattering process. The combination of these different scattering processes gives rise to multiple channel modes (see Eqs. (\ref{ckkco1}) and (\ref{magetoc})). The competition between these modes quantitatively reveals the crossover from positive to negative magnetoconductivity when crossing the VHS energy (refer to Ref. \cite{supp} for detailed analysis).

We investigate the local properties surrounding the VHS, where the dispersion relation is described by a momentum-dependent quadratic equation: \begin{equation}
E=b+\frac{\gamma^2}{2b}k_\parallel^2-v_0k_z^2.\label{lowen}
\end{equation}
In this equation, the opposite signs of the $k_\parallel$ and $k_z$ terms generate a band structure resembling a flat band, resulting in a significant increase in the density of states (see supplemental material \cite{supp}). This heightened density of states amplifies electron scattering due to the larger number of available states for scattering processes near the VHS.
In close proximity to the VHS, the electron behavior is primarily governed by Eq. (\ref{lowen}). Consequently, scattering path 1 emerges as the dominant process in the presence of impurity potentials. Its contribution to the magnetoconductivity exceeds that of path 2, leading to positive magnetoconductivity. However, as we move further away from the VHS, the influence of the VHS diminishes, and the contribution from path 2 becomes the main determinant of the magnetoconductivity, resulting in a negative value (refer to supplemental material \cite{supp}), which is consistent with the observations in Figure \ref{fig3}(a).

%**********************************************************************************************************************************
\begin{figure}[tbp]
\centering
\includegraphics[width=0.45\textwidth]{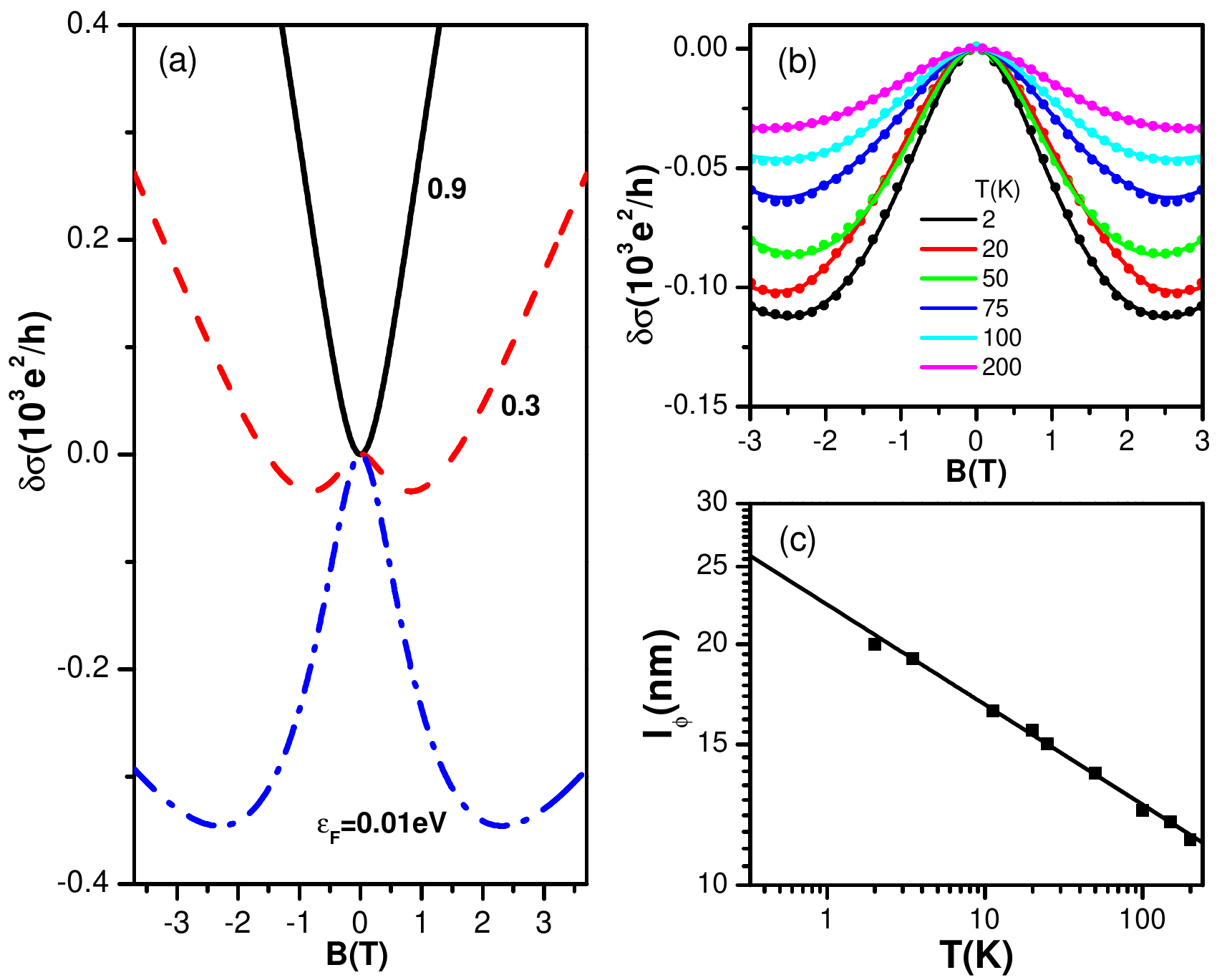}%0.38
\caption{Magnetoconductivity versus the magnetic field for different Fermi energy $\varepsilon_F$ at $l_\phi=100$ nm in (a); and for different temperature $T$ at $\varepsilon_F=0.015$ eV in (b). In (b), the filled symbols are experimental data extracted from Ref. \cite{schumannprb2017}, and the solid lines are the theoretical fittings based on Eq. (\ref{magetoc}). (c) Temperature dependence of the fitted phase coherence length $l_\phi$ (filled circles), where the solid line comes from the relation $l_\phi=CT^{-0.125}$ with $C=22.42$. The other parameters are the same as those for Figure \ref{fig3}.% $\gamma=1.0$ eV$\cdot$nm, $v_0=0.6$ eV$\cdot$nm$^2$, %$b=0.9$eV, and %$l_e=10$nm.
\label{fig4}}
\end{figure}
%************************************************************************************************************************************
We investigated the variation of magnetoconductivity with magnetic field, as depicted in Figure \ref{fig4}. Our observations revealed a magnetic field-driven crossover from negative to positive magnetoconductivity at an appropriate Fermi energy value (indicated by the red line in Figure \ref{fig4}(a)), corresponding to the staggered intersection of magnetoconductivity curves shown in Figure \ref{fig3}(a). This crossover behavior can be eliminated by adjusting the Fermi energy(see Figure \ref{fig4}(a)). While a similar crossover feature has been observed in single Dirac node systems with magnetic doping or intervalley scattering \cite{haizhou2013, luprl2011}, it is not solely determined by the Fermi energy.
We compared the magnetic field dependence of magnetoconductivity at different temperatures with experimental data for the Cd$_3$As$_2$ thin film, which has a thickness of 370 nm \cite{schumannprb2017}. As the temperature increases, we observed a distortion of the W-shaped structure of the magnetoconductivity versus magnetic field (see Figure \ref{fig4}(b)), which aligns well with the experimental results. Based on this theoretical fitting, we established the relationship between the phase coherence length and temperature, represented by black squares in Figure \ref{fig4}(c). We discovered that this relationship follows the equation $l_\phi = C T^{-p/2}$ with the parameters $C=22.42$ and $p=0.125$ (solid line in Figure \ref{fig4}(c)), derived by considering the influence of electron-electron interaction\cite{leermp1985}. This finding suggests that the phase coherence length in the Cd$_3$As$_2$ thin film is governed by electron-electron interaction, and provides further evidence for the role of VHS in magnetoconductivity.
%**********************************************************************************************************************************
\begin{figure}[tbp]
\centering
\includegraphics[width=0.48\textwidth]{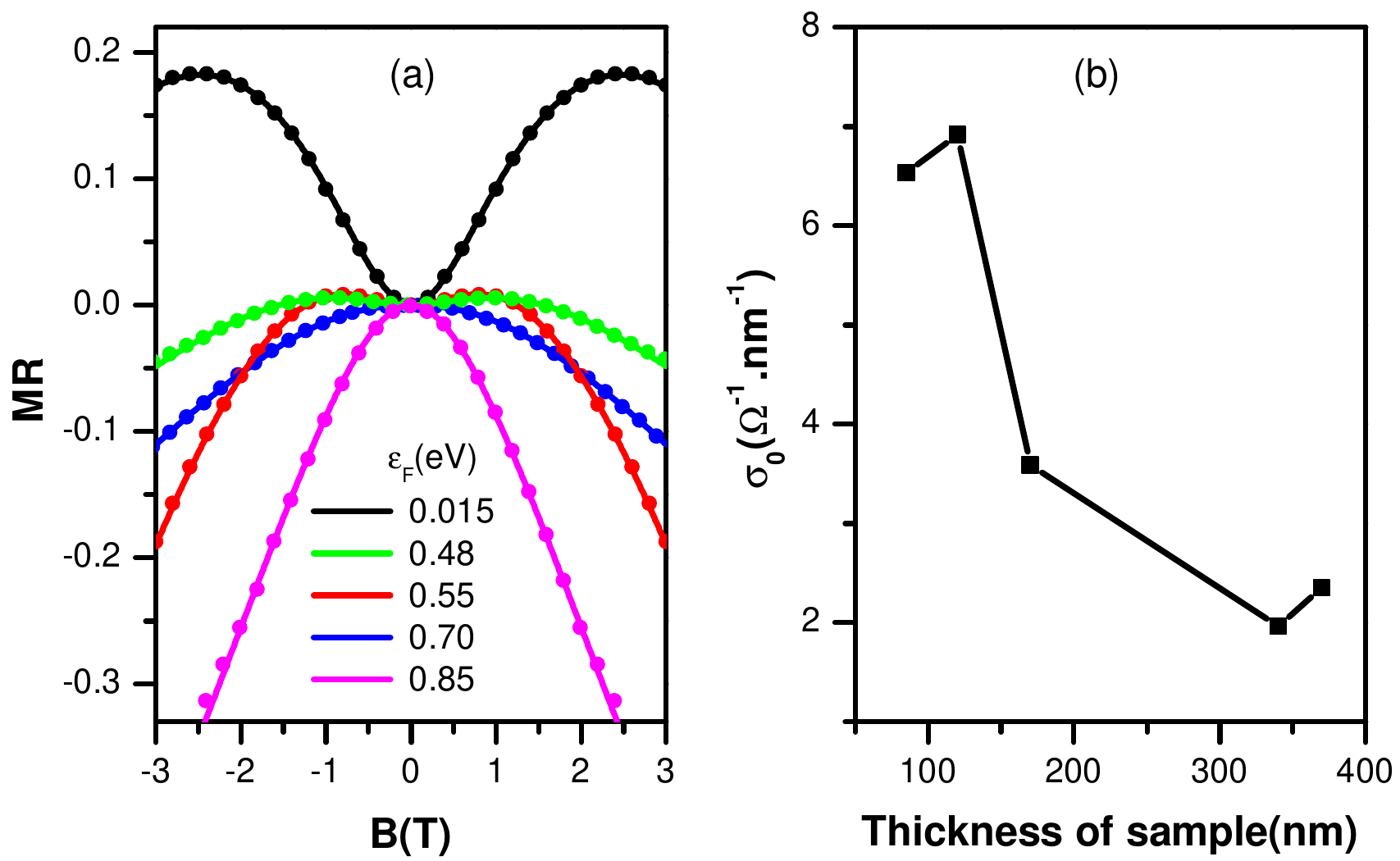}%0.38
\caption{ (a)
Theoretical fitting for the  magnetoresistance of the samples Cd$_3$As$_{2}$ at $T=2$K, where the solid curves are the theoretical results for different Fermi energies based on Eq.(\ref{magetoc}), and the filled circles are the experimental data for the thickness $370$ nm (Black),$85$nm(Green), $340$ nm (Red), $120$ nm(Blue) and $170$nm(Magenta), extracted from Ref. \cite{schumannprb2017}. The fitted zero-field conductivity as a function of the sample thickness is shown in Figure (b). %$\sigma_0=2.35\Omega^{-1}(nm)^{-1}$(black), $1.96\Omega^{-1}(nm)^{-1}$(red), $6.92\Omega^{-1}(nm)^{-1}$(blue). %and %$\varepsilon_F=0.015eV$(black), $0.55eV$(red), $0.7 %eV$(blue).
The other parameters are taken the same as those of Figure \ref{fig3}.\label{fig5}}
\end{figure}
%************************************************************************************************************************************

\textit{Magnetoresistance.---} The magnetoresistance is defined as
$
\text{MR}=[R(B)-R(0)]/R(0)
$, where $R(B)$  represents the resistance in the presence of the external magnetic field. By substituting $R(B)=S/[\sigma_0+\delta\sigma(B)]L$, where $\sigma_0$ is the conductivity at zero magnetic field, $S$ and $L$ are the cross-sectional area and length of the sample, we obtain
\begin{equation}
\text{MR}=-\frac{\delta\sigma(B)}{\sigma_0+\delta\sigma(B)}.
\end{equation}

In Figure \ref{fig5}, the magnetoresistance is plotted as a function of the magnetic field for different Fermi energies. Notably, a distinct V-shaped notch is observed near $B=0$ for small Fermi energies. As the Fermi energy increases, the amplitude of the notch gradually decreases. Moreover, when the Fermi energy approaches the VHS, the magnetoresistance undergoes a transition from positive to negative, with the effect becoming more prominent for larger magnetic field values and extending towards $B=0$. We conducted a fitting analysis using experimental data obtained from Cd$_3$As$_2$ thin film samples with varying thicknesses\cite{schumannprb2017}, as shown in Figure \ref{fig5}. The fitting results exhibit excellent agreement with the experimental data. Furthermore, our fitting analysis revealed a nonmonotonic dependence of the zero-field conductivity on the sample thickness, as depicted in Figure \ref{fig5}(b), which is consistent with experimental findings reported by Schumann, \emph{et al.} \cite{schumannprb2017}. This consistency, combined with the excellent fitting results, strongly supports the assertion that the VHS is a crucial factor contributing to the occurrence of NMR.

\textit{Discussions and conclusions.---}
 A recent experiment has unveiled the occurrence of NMR in thin films of the Dirac semimetal Cd$_3$As$_2$ \cite{schumannprb2017}. However, the explanation for the dependence of this phenomenon on film thickness remains incomplete.
Cd$_3$As$_2$ exhibits Dirac nodes that are slightly close in proximity \cite{tlnm2015,zhaosr2016,schumannprb2017}, leading to a relatively low energy VHS. In thin systems, the confinement effect lifts the Fermi energy \cite{dymnikovpss2011}, causing it to approach the VHS energy of the Dirac semimetal Cd$_3$As$_2$. Notably, when the Fermi energy crosses the VHS, a Lifshitz phase transition occurs in the Fermi surface \cite{lifshitz1960}. In such cases, the role of the VHS becomes significant and cannot be ignored. The two-node model employed in this study effectively encompasses these fundamental physical characteristics. Through this model, we demonstrate that the VHS increases the density of states, thereby enhancing backscattering due to the greater number of available states for scattering processes. This boosted backscattering results in increased magnetoconductivity and NMR, which agrees well with the experimental results, evidencing the key role of the VHS in magnetotransport.

%\begin{acknowledgments}

\textit{Acknowledgments.-} We thank Professor Zheng-Yu Weng for the valuable suggestions, Professors Susanne Stemmer, Baigeng Wang, and Fengqi Song for providing the original experimental data. The work is supported by the Natural Science Foundation
of Hunan Province, China (Grant No. 2023JJ30005), and the Construct Program of the Key Discipline of Hunan Province.
ZGZ is supported in part by the NSFC (Grants No. 11974348 and No. 11834014), and the Strategic Priority Research Program of CAS (Grants No. XDB28000000, and No. XDB33000000), the National Key R\&D Program of China (Grant No. No. 2022YFA1402802), the Training Program of Major Research plan of the National Natural Science Foundation of China (Grant No. 92165105), and CAS Project for Young Scientists in Basic ResearchGrant No. YSBR-057.

%\end{acknowledgments}


\begin{thebibliography}{19}


\bibitem{pippard}A. B. Pippard, \emph{Magnetoresistance in Metals}, Cambridge
University Press, Cambridge, 1989.

\bibitem{beenakker} C. W. J. Beenakker and H. van Houten, Solid State Phys.
\textbf{44}, 1(1991).

\bibitem{lin2011} E. K. Lin and M. Cicerone, Adv. Mater. \textbf{23}, 317(2011).

\bibitem{mn2007} M. Nishioka, Y. B. Lee, and A. M. Goldman, Y. Xia, and C. Daniel Frisbie, Appl. Phys. Lett. \textbf{91}, 092117 (2007).

\bibitem{bloom2007} F. L. Bloom, W. Wagemans, M. Kemerink, and B. Koopmans
Phys. Rev. Lett. \textbf{99}, 257201(2007).




\bibitem{kimprl2013} H. J. Kim, K. S. Kim, J. F. Wang,  M. Sasaki, N. Satoh, A. Ohnishi,  M. Kitaura,  M. Yang, and L. Li, Phys. Rev. Lett. \textbf{111}, 246603 (2013).
\bibitem{tlnm2015} T. Liang, Q. Gibson, M. N. Ali, M. Liu, R. J. Cava, and N. P. Ong,
Nat. Mater. \textbf{14}, 280 (2015).




\bibitem{zhaosr2016} B. Zhao, P. Cheng, H. Pan, S. Zhang, B. Wang, G. Wang, F. Xiu, and F. Song, Sci. Rep. \textbf{6}, 22377 (2016).

\bibitem{schumannprb2017} T. Schumann, M. Goyal, D. A. Kealhofer, and S. Stemmer, Phys. Rev. B \textbf{95}, 241113(R) (2017).




\bibitem{zklsci2014}Z. K. Liu, B. Zhou, Y. Zhang, Z. J. Wang, H. M. Weng, D. Prabhakaran, S. K. Mo, Z. X. Shen, Z. Fang, X. Dai, Z. Hussain, and Y. L. Chen,
Science \textbf{343}, 864 (2014).


\bibitem{xiongsci2015} J. Xiong, S. K. Kushwaha, T. Liang, J. W. Krizan, M.
Hirschberger, W. Wang, R. J. Cava, and N. P. Ong, Science
\textbf{350}, 413 (2015).

\bibitem{linp2016} Q. Li, D. E. Kharzeev, C. Zhang, Y. Huang, I. Pletikosie, A. V. Fedorov, R. D. Zhong, J. A. Schneeloch, G. D. Gu, and T. Valla,
Nat. Phys. \textbf{12}, 550 (2016).









\bibitem{wanprb2011}
 X. Wan, A. M. Turner, A. Vishwanath, and S. Y. Savrasov,
Phys. Rev. B \textbf{83}, 205101 (2011).

\bibitem{xuprl2011} G. Xu, H. Weng, Z. Wang, X. Dai, and Z. Fang,  Phys. Rev. Lett. \textbf{107},
186806 (2011).

\bibitem{lvprx2015}B. Q. Lv, H. M. Weng, B. B. Fu, X. P. Wang, H. Miao, J. Ma, P. Richard, X. C. Huang, L. X. Zhao, G. F. Chen, Z. Fang, X. Dai, T. Qian, and H. Ding,
Phys. Rev. X \textbf{5}, 031013 (2015).


\bibitem{aasnat2015} A. A. Soluyanov, D. Gresch, Z. Wang, Q. Wu, M. Troyer, X. Dai, and B. A. Bernevig, Nature \textbf{527}, 495 (2015).

\bibitem{hxwprl2020}  H. X. Wang, Z. K. Lin, B. Jiang, G. Y. Guo, and J. H. Jiang,
Phys. Rev. Lett. \textbf{125}, 146401 (2020).


\bibitem{xiaormp2000} D. Xiao, M. C. Chang, and Q. Niu, Rev. Mod. Phys. \textbf{82}, 1959(2010).


\bibitem{aazprb2012} A. A. Zyuzin and A. A. Burkov, Phys. Rev. B \textbf{86}, 115133
(2012).

\bibitem{dtsprl2012} D. T. Son and N. Yamamoto, Phys.
Rev. Lett. \textbf{109}, 181602 (2012).
\bibitem{dtsprb2013} D. T. Son and B. Z. Spivak, Phys. Rev. B \textbf{88}, 104412
(2013).
\bibitem{ashby2014} P. E. C. Ashby and J. P. Carbotte, Phys. Rev. B \textbf{89}, 245121
(2014).










\bibitem{vaprb2012} V. Aji, Phys. Rev. B \textbf{85}, 241101(R) (2012).








\bibitem{aabprl2014} A. A. Burkov, Phys. Rev. Lett. \textbf{113}, 247203 (2014).

\bibitem{xhprx2015} X. Huang, L. Zhao, Y. Long, P. Wang, D. Chen, Z. Yang, H.
Liang, M. Xue, H. Weng, Z. Fang, X. Dai, and G. Chen,
Phys. Rev. X \textbf{5}, 031023 (2015).

\bibitem{daiprl2017} X. Dai, Z. Z. Du, and H. Z. Lu, Phys. Rev. Lett. \textbf{119}, 166601(2017).

\bibitem{reisnjp2016} R. D. dos Reis, M. O. Ajeesh, N. Kumar, F. Arnold, C. Shekhar, M. Naumann, M. Schmidt, M. Nicklas, and E. Hassinger, New J. Phys. \textbf{18}, 085006 (2016).

\bibitem{wangnr2012} J.Wang, H. Li, C. Chang, K. He, \emph{et al.}, %J. S. Lee, H. Lu, Y. Sun, X.
%Ma, N. Samarth, S. Shen \emph{et al.},
Nano Res. \textbf{5}, 739 (2012).



\bibitem{mutchsci2019} J. Mutch, W. C. Chen, P. Went, T. Qian, I. Z. Wilson, A.
Andreev, C. C. Chen, and J. H. Chu, Science Advances \textbf{5}, eaav9771(2019).

\bibitem{fuprl2019} B. Fu, H. W. Wang, and S. Q. Shen, Phys. Rev. Lett. \textbf{122}, 246601 (2019).

\bibitem{luprb2015} H. Z. Lu, S. B. Zhang, and S. Q. Shen, Phys. Rev. B \textbf{92}, 045203
(2015).
\bibitem{changprb2015} M. C. Chang and M. F. Yang, Phys. Rev. B \textbf{91}, 115203 (2015).

\bibitem{cortijoprb2016} A. Cortijo, Phys. Rev. B \textbf{94}, 235123 (2016).

\bibitem{lifshitz1960}I. M. Lifshitz, Sov. Phys. JETP \textbf{11}, 1130 (1960).

\bibitem{dingprb2021} K. H. Ding, Z. G. Zhu, Y. L. Hu, and G. Su, Phys. Rev. B \textbf{104}, 155135 (2021).

\bibitem{xuprb2015} S. Y. Xu, C. Liu, I. Belopolski, S. K. Kushwaha, R. Sankar, J. W. Krizan, \emph{et al.}, %T.-R. Chang, C. M. Polley, J. Adell, T. Balasubramanian, K. %   Miyamoto, N. Alidoust, G. Bian, M. Neupane, H.-T. Jeng, C.-Y. Huang, W.-F. Tsai, T. Okuda, A. Bansil, F. C. Chou, R. J. Cava, H. Lin, and M. Z. Hasan,
Phys. Rev. B \textbf{92}, 075115 (2015).



\bibitem{chenprb2015} R. Y. Chen, S. J. Zhang, J. A. Schneeloch, C. Zhang, Q. Li, G. D. Gu, and N. L. Wang, Phys. Rev. B \textbf{92}, 075107 (2015).




\bibitem{Akkermansbook} E. Akkermans and G. Montambaux, \emph{Mesoscopic Physics
of Electrons and Photons} (Cambridge University Press,
Cambridge, England, 2007).

\bibitem{Sinitsynprb2007} N. A. Sinitsyn, A. H. MacDonald, T. Jungwirth, V. K. Dugaev, and J. Sinova,
Phys. Rev. B \textbf{75}, 045315 (2007).

\bibitem{hikami1980} S. Hikami, A. I. Larkin, and Y. Nagaoka, Prog. Theor.
Phys. \textbf{63}, 707 (1980).

\bibitem{bruusbook} H. Bruus and K. Flesberg, \emph{Many-body Quantum Theory
in Condensed Matter Physics} (Oxford University Press,
2004).

\bibitem{altshuler1980} B. L. Altshuler, D. Khmelnitzkii, A. I. Larkin, and P. A.
Lee, Phys. Rev. B \textbf{22}, 5142 (1980).

\bibitem{mccann06} E. McCann, K. Kechedzhi, V. I. Fal'ko, H. Suzuura, T. Ando, and B. L. Altshuler,
Phys. Rev. Lett. \textbf{97}, 146805 (2006).

\bibitem{luprl2011} H. Z. Lu, J. Shi, and S. Q. Shen, Phys. Rev. Lett. \textbf{107}, 076801 (2011).


 \bibitem{tkachovprb2011}G. Tkachov and E. M. Hankiewicz, Phys. Rev. B \textbf{84}, 035444 (2011).



\bibitem{supp} Supplemental material.% at ¡¤ ¡¤ ¡¤ for the detailed calculations
%on the Cooperon, the renormalization of the velocity, the quan-
%tum conductivity correction, and the magnetoconductivity in
%the cases of the even and odd $\Delta$.




\bibitem{chenprl2019} W. Chen, H. Z. Lu, and O. Zilberberg, Phys. Rev. Lett. \textbf{122}, 196603 (2019).

\bibitem{daiprb2016} X. Dai, H. Z. Lu, S. Q. Shen, and H. Yao, Phys. Rev. B \textbf{93}, 161110(R) (2016).

\bibitem{suzuura02}H. Suzuura and T. Ando, Phys. Rev. Lett. \textbf{89}, 266603 (2002).

\bibitem{haizhou2013} H. Z. Lu, W. Yao, D. Xiao, and S. Q. Shen,
Phys. Rev. Lett. \textbf{110}, 016806 (2013).

\bibitem{leermp1985} P. A. Lee and T. V. Ramakrishnan, Rev. Mov. Phys. \textbf{57}, 287(1985).

\bibitem{dymnikovpss2011} V. D. Dymnikov,
Phys. Sol. Stat.\textbf{ 53}, 901 (2011).

%\bibitem{youngprl} S. M. Young, S. Zaheer, J. C. Y. Teo, C. L. Kane, E. J. Mele, and A. M. Rappe,
%Phys. Rev. Lett. \textbf{108}, 140405 (2012).

%\bibitem{mafiesprb2012} J. L. Ma$\tilde{n}$es, Phys. Rev. B \textbf{85}, 155118 (2012).

%\bibitem{wangprb2013} Z. Wang, H. Weng, Q. Wu, X. Dai and Z. Fang, Phys. Rev. B \textbf{88}, 125427 (2013).

%\bibitem{wangapx2017} S. Wang, B. C. Lin, A. Q. Wang, D. P. Yu, and Z. M. Liao,
%Adv. Phys. X \textbf{2}, 518 (2017).























































































\end{thebibliography}
 \end{document}